# Quantum Private Comparison: A Review

Wenjie Liu[1,2], Chao Liu[2], Haibin Wang[2] and Tingting Jia[2]

[1]Jiangsu Engineering Center of Network Monitoring, [2]School of Computer and Software, Nanjing University of Information Science and Technology, Nanjing, China

### Abstract

As an important branch of quantum secure multiparty computation, quantum private comparison (QPC) has attracted more and more attention recently. In this paper, according to the quantum implementation mechanism that these protocols used, we divide these protocols into three categories: The quantum cryptography QPC, the superdense coding QPC, and the entanglement swapping QPC. And then, a more in-depth analysis on the research progress, design idea, and substantive characteristics of corresponding QPC categories is carried out, respectively. Finally, the applications of QPC and quantum secure multi-party computation issues are discussed and, in addition, three possible research mainstream directions are pointed out.

## 1. Introduction

Secure multi-party computation (SMC) is a basic topic in the distributed computation field, which allows a group of mutually distrustful players to perform correct, distributed computations without leaking their respective secret inputs under the sole assumption that some of them will follow the protocol honestly. The Millionaire problem introduced by Yao [1] was the origin of secure two-party computation, in which two millionaires wish to know who is richer without revealing the precise amount of their fortunes. Based on Yao's millionaires problem, Boudot *et al.* [2] subsequently proposed a protocol to decide whether two millionaires are equally rich. However, Lo [3] pointed out that the equality function cannot be securely evaluated with a two-party scenario. Therefore, some additional assumptions (e.g., a semi-honest third party) should be considered to reach the goal of private comparison. SMC problem has been studied extensively in the classical setting, but the security is based on computational complexity assumptions, so it is conditionally secure. With the improvement of computation power and presentation of novel algorithms, especially the appearance of quantum computer and quantum algorithms, some classic NP-complete problems have been being constantly broken; therefore, those protocols based on computational complexity encounter serious challenges. Different from classical counterpart, secure computation in the quantum mechanism can attain unconditional security because the security is ensured by some physical principles, such as the Heisenberg uncertainty principle and the quantum no-cloning theorem. Under this background, making use of quantum mechanical effect [4] to solve SMC problem has been attracting more and more attention.

As an important branch of quantum secure multi-party computation (QSMC), quantum private comparison (QPC) [5-16] has been extensively discussed and studied, it can be applied extensively in many application fields, including private bidding and auctions, secret ballot elections, commercial business, identification in a number of scenarios, and so on. Being a QPC protocol, it must ensure:

- Fairness: The protocol is fair, which means that one party knows the sound result of comparison result if and only if the other parties know the sound result
- Security: Outside parties cannot learn any information about players' private inputs, and they cannot deduce their secret inputs from the comparison result, moreover, one player cannot know the other's secret input
- Efficiency: With the help of the TP, the protocol should save more quantum and classical resources in contrast to the case without the TP.

In most previous QPC protocols, TP is assumed to be semi-honest, i.e., he/she will help the parties accomplish the comparison and executes the protocol loyally, but takes a record of all intermediate computations and might try to steal the information from the record. The point to be mentioned is that TP will not be corrupted by an outside eavesdropper.

In QPC protocols, "the private information is equal or not" is usually taken as the comparison result, the detailed description of a two-party QPC problem is as follows:

Input: Alice has a private information *x*, Bob has a private information *y*. The binary representations of *x* and *y* are $(x_0, x_1, ..., x_{N-1})$, $(y_0, y_1, ..., y_{N-1})$, where $x_i, y_i \in \{0,1\}$; $x = \sum_{i=0}^{N-1} x_i 2^i$, $y = \sum_{i=0}^{N-1} y_i 2^i$, $2^{N-1} \leq \max\{x, y\} \leq 2^N$

Output: $x = y$ or $x \neq y$.

According to quantum mechanism that the QPC protocols utilize, we can divide them into three categories: The quantum cryptography QPC [5,6], the superdense coding QPC [7-13], and the entanglement swapping QPC [14-16]. Using quantum cryptography to deal with private comparison is one of the most common ways, which uses quantum states to transfer encryption keys, and then takes these keys to encrypt private information. The superdense coding QPC introduces some unitary operators to encode private information into quantum states, and deduces the private comparison results by measuring the final quantum state. Different from the above two, the advantage of entanglement swapping QPC is that all particles carrying the secret messages only need the one-way transmission, so it is easier to be implemented in the quantum one-way computer.

The structure of this paper is organized as follows, in Section 2, the research progress, design idea and substantive characteristics of quantum cryptography QPC are elaborated and analyzed in depth; the same elaboration and analysis on the superdense coding QPC and the entanglement swapping QPC are done in Section 3, Section 4 respectively. At last, the applications of QPC and other QSMC are discussed, and three research directions are pointed out as well in Section 5.

## 2. Quantum Cryptography QPC

Using quantum cryptography to design QPC protocols may be a usual way to be chosen. In 2010, Chen *et al.* [5] proposed a quantum cryptography QPC protocol via triplet GHZ (Greenberger-Horne-Zeilinger) states. However, Lin *et al.* [17] pointed out that one can retrieve other's secret information by means of intercept-resend attack [18], because the positions of checking particles or the measurement basis are determined by the participants in the eavesdropping check phase. Moreover, they gave two solutions to avoid this attack, i.e., they let the third-party determine the positions and the measurement basis. In order to improve qubit efficiency, in 2012, Tseng *et al.* [6] proposed quantum cryptography QPC protocol using Bell state, which is easier to be achieved in the physical experiment, and its qubit efficiency is near 50%. However, Yang *et al.* [19] pointed out that if the third-party is disloyal, Tseng *et al.* protocol has an obvious security flaw: TP can know the two players' private information. In order to make up this flaw, Yang suggested that it should add a loyal checking procedure for the third-party after transmission security detection.

Taking the triplet GHZ states as quantum resource, let us briefly introduce how to realize the quantum cryptography QPC [Figure 1].

Step 1. Suppose that two players Alice, Bob and the third-party TP agree that $|+\rangle$ ($|-\rangle$) represents information '0'('1'). TP prepares *N* triplet GHZ states as follows:

$$|\Psi\rangle = \frac{1}{\sqrt{2}}(|000\rangle + |111\rangle)$$
$$= \frac{1}{2\sqrt{2}}(|+++\rangle + |+--\rangle + |-+-\rangle + |--+\rangle), \quad (1)$$

And then, he/she sends the first (second) particles sequence to Alice (Bob) and retains the third particles sequence.

Step 2. In order to ensure the quantum channel security, Alice and Bob need to perform an eavesdropping check procedure. After that, they make single-particle measurement in the *X*-basis $\{|+\rangle, |-\rangle\}$ on the remaining particles, and obtain the encryption key $K_A$, $K_B$, respectively. Now, Alice and Bob can encrypt their secret messages as follows, $C_A = x \oplus K_A$, $C_B = y \oplus K_B$, and then compute $C = C_A \oplus C_B$. Finally, *C* is sent to TP.

Step 3. If the *i*th value in *C* is '1', TP performs unitary operation $\sigma_Z = (|0\rangle\langle 0| - |1\rangle\langle 1|)$ on the *i*th qubit of the TP's retaining sequence. If not, TP does nothing.

Step 4. TP performs the *X*-basis measurement on the particles on his/her hand, if any $|-\rangle$ exists in the measuring result, we know $x \neq y$; otherwise, $x = y$.

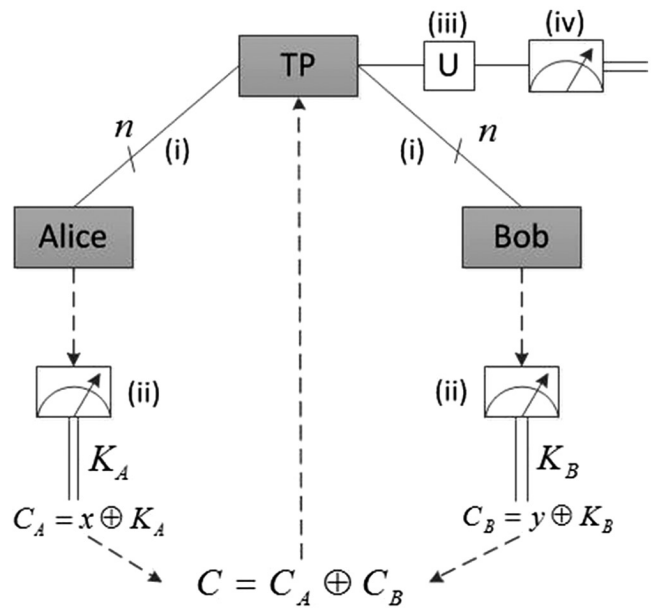

**Figure 1:** Process of the quantum cryptography QPC.

The essence of quantum encryption QPC is to use quantum state to transfer encryption keys (the phase is equivalent to the QKD protocol [20,21]), then utilize these keys to encrypt secret messages, finally make some exclusive-OR operations to judge whether the secret message is equal or not. Since QKD protocol has been proven to be unconditionally secure [22,23], it means that other parties can't gain any information of the secret message from encrypted text. Compared with other types of QPC, the quantum cryptography QPC is simpler and easier in the physical implementation.

## 3. Superdense Coding QPC

Quantum superdense coding [24] is an important application in quantum computation and quantum information fields. In contrast to the classic coding, the advantage is the high efficiency: It can transmit two classical bits of information by only sending one quantum bit. In 2009, Yang *et al.* [7] proposed the pioneering superdense coding QPC protocol using Bell states, which utilizes four unitary operations (I, $\sigma_x$, $\sigma_z$, $i\sigma_y$) to encode information. Afterwards, for the sake of saving the qubit, Yang *et al.* [8] presented a novel QPC protocol employing single photons. In the protocol, they utilize the $I/\sigma_x$ operations to encode information, and use decoy photons to detect outsider's cheating. In Ref. [9], Liu *et al.* proposed another single photon QPC protocol by introducing QKD and the Hash function to encrypt secret messages. In order to avoid inserting repeatedly the decoy photons and performing the step-by-step channel detection, they use the collective detection to reduce the photon consumption and simplify the protocol steps. In addition, some researches have been done on non-maximally entangled state QPC. In 2011, Liu *et al.* [10] proposed a superdense coding QPC protocol based on triplet W states. However, Li *et al.* [11] pointed out that one player can estimate the other's private bit successfully with probability 2/3, rather than 1/2. And then, they presented a secure protocol using Bell states instead of W states, but it is a pity that they did not find an effective solution with W states. Recently, in order to improve the efficiency, Jia *et al.* [12] put forward a QPC protocol based on four-particle χ-Type states. Due to no need to publish any message after operations encoding, the efficiency of the protocol is equal to 100%. At the same time, Liu *et al.* [13] proposed another QPC protocol utilizing four-particle χ-Type states. Different from Jia's protocol, it only needs to prepare one type of χ-type state, and the amount of χ-type states is only $N$, which saves more quantum resources.

The χ-Type is defined as follows:

$$|\chi^{00}\rangle_{1234} = \frac{1}{2}(|00\rangle|\Phi^+\rangle + |11\rangle|\Phi^-\rangle - |01\rangle|\Psi^-\rangle + |10\rangle|\Psi^+\rangle)_{1234}, \quad (2)$$

where,

$$|\Phi^\pm\rangle = \frac{1}{\sqrt{2}}(|00\rangle \pm |11\rangle), \quad |\Psi^\pm\rangle = \frac{1}{\sqrt{2}}(|01\rangle \pm |10\rangle). \quad (3)$$

Let us give a brief description on the superdense coding QPC protocol using χ-Type states, [Figure 2].

Step 1. Suppose that Alice and Bob agree $|0\rangle$, $|\Phi^\pm\rangle$ represent information '0'; $|1\rangle$, $|\Psi^\pm\rangle$ represent information '1'. TP prepares $N$ χ-Type states and sends the first (second) particles sequence to Alice (Bob), he/she retains the third, four particles sequence in hand.

Step 2. After Alice and Bob received sequences, they perform eavesdropping check. After that, Alice and Bob perform Pauli operation on the received particles according to their secret message. If secret message is '1', $\sigma_x = (|0\rangle\langle 1| + |1\rangle\langle 0|)$ operation is utilized; otherwise $I = (|0\rangle\langle 0| + |1\rangle\langle 1|)$ operation is utilized. And then Alice and Bob use the Z-basis $\{|0\rangle, |1\rangle\}$ to measure each particle and obtain $R_a$ and $R_b$, respectively. They calculate $R = R_a \oplus R_b$ and send $R$ to TP.

Step 3. TP makes the Bell measurement on the third, four particles sequence, if the result is $|\Phi^\pm\rangle$, $R_t = 0$; if the result is $|\Psi^\pm\rangle$, $R_t = 1$. After that, TP calculates the bit-wise exclusive-OR operation between $R$ and $R_t$, if the result is '0', then $x = y$; otherwise $x \neq y$.

In the superdence coding QPC, the core issue is how to select the appropriate unitary operator according to the players' secret message. The chosen unitary operator will be used to encode these secret messages into quantum states, and we can get comparison result by measuring these quantum states. Compared with the quantum cryptography QPC, the superdense coding QPC has a high efficiency, and it directly uses the quantum channel to transmit the secret messages, which greatly exerts the unconditional security of quantum cryptography. Therefore, to some extent, the superdense coding QPC is a special application of quantum direct communication (QDC) [25-27] in the private information comparison field.

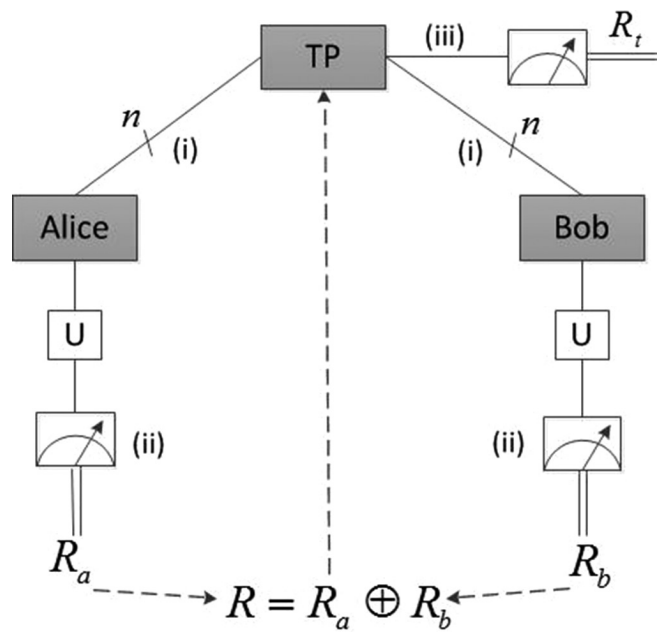

**Figure 2:** Process of the superdense coding QPC.

## 4. Entanglement Swapping QPC

Entanglement swapping [28,29] plays a very important role in quantum information. Through entanglement swapping, one can entangle particles that do not even share any common past. Suppose the initial state is $|\Phi^+\rangle_{12} \otimes |\Phi^+\rangle_{34}$, here $|\Phi^+\rangle$ is defined in Equation (2). After the Bell measurements on the pair of particle 1 and 3, the pair of photon 2 and 4 will be entangled.

$$|\Phi^+\rangle_{12} \otimes |\Phi^+\rangle_{34} = \frac{1}{2}(|\Phi^+\rangle_{13}|\Phi^+\rangle_{24} + |\Phi^-\rangle_{13}|\Phi^-\rangle_{24}$$
$$+ |\Psi^+\rangle_{13}|\Psi^+\rangle_{24} + |\Psi^-\rangle_{13}|\Psi^-\rangle_{24}) \quad (4)$$

When the outcome of Bell measurement of particle 1 and 3 is $|\Phi^+\rangle$, the Bell state of remaining photons must be $|\Phi^+\rangle$.

In 2012, Liu *et al.* [14] presented a QPC protocol based on four-particle χ-Type states entanglement swapping. Recently, Liu *et al.* [15,16] proposed QPC protocols based on Bell states and triplet GHZ states, respectively. Taking Bell states as an example, the brief procedures of an entanglement swapping QPC protocol can be included as follows. [Figure 3].

Step 1. Suppose that Alice and Bob agree that $|\Phi^+\rangle$, $|\Phi^-\rangle$, $|\Psi^+\rangle$, $|\Psi^-\rangle$ represents information '00', '10', '10', '11', respectively. Alice, Bob and TP prepare an order [N/2] EPR pairs of $|\Phi^+\rangle$, and take particle 1 (2) to form their own first (second) particles sequence $S_1^A$ ($S_2^A$), $S_1^B$ ($S_2^B$) and $S_1^T$ ($S_2^T$), respectively.

Step 2. Alice and TP prepare an order N EPR pairs sequence of $|\Phi^+\rangle$ once again, and these sequences are used to check eavesdroppers. Alice (TP) inserts the first and second particles of N EPR pairs into sequences $S_1^A$ ($S_1^T$) and $S_2^A$ ($S_2^T$), respectively. Afterward, they exchange the new $S_2^A$, $S_2^T$ sequences.

Step 3. After received all particles, Alice and TP perform eavesdropper checking. If they confirm there is no eavesdropper, Alice performs Bell-basis measurement on two corresponding particles and gets $R_j^A$ from measurement result. If measurement result is $|\Phi^+\rangle(|\Phi^-\rangle, |\Psi^+\rangle, |\Psi^-\rangle)$, then $R_j^A$ is '00'('01', '10', '11'). Then, the two corresponding particles in TP's sequence are collapsed into one of four Bell states.

Step 4. Bob and TP perform the similar process as Step 2 and 3, respectively. Finally, They get $R_j^B$ and $R_j^C$, here $R_j^C = (r_j^{C1} r_j^{C2})$.

Step 5. Alice and Bob calculate $R_j = (R_j^A \oplus x_j) \oplus (R_j^B \oplus y_j) = (r_j^1 r_j^2)$ $(1 \leq j \leq N/2)$, and send the result $R_j$ to TP. TP calculates $R = \sum_{j=1}^{\lceil N/2 \rceil} ((r_j^1 \oplus r_j^{C1}) + (r_j^2 \oplus r_j^{C2}))$, and send R to Alice and Bob, if R = 0, then x = y; otherwise x ≠ y.

In the entanglement swapping QPC protocols, the secret messages can be encrypted in quantum cryptography or superdense coding method, and the different key point is that this kind of protocols utilize quantum entanglement swapping characteristic to perform the comparison task. Speaking in detail, it can product nonlocal entanglement correlation by measuring the remote unrelated particles, and this nonlocal correlation will be used to realize the private message comparison. The efficiency of entanglement swapping QPC may not be ideal, but it plays a very important role in some certain conditions, e.g., there is no available quantum channel between the communication players. Moreover, it has a significant advantage: The particles carrying the secret messages only need one-way transmission, so it is feasible to implement in the one-way quantum computer [30,31].

## 5. Discussions and Perspectives

The QPC of equal information is only the simplest form of QPC. With the deepening research of QPC, multi-party quantum private comparison (MQPC) and quantum private comparison in size (SQPC) have been gradually aroused the researchers' attention. Taking MQPC as an example, it is more versatile and practically valuable. Obviously, you can utilize the above two-party comparison protocols to realize multi-party comparison. However, with the increase of the players' scale, the efficiency will drop quickly, in the worst case, the number of comparison is $n(n-1)/2$, so the protocol for multi-party comparison needs to be reconsidered and redesigned. Recently, a research group has proposed a MQPC protocol based on multi-particles GHZ states [32], which allows *n* players to compare their own private information and get the result within one comparison execution. Besides, for SQPC, it is significant in some practical application. Due to the comparison, result may be greater or less or equal; it is difficult to realize by using the previous two-dimensional quantum state. As we all known, the quantum particle has the natural multi-dimensional characteristic and the high-dimensional quantum state shows greater capacity than two-dimensional quantum state. Taking Bell state as an example, the capacity of *d*-dimensional Bell state is $\log_2 d$ times than its two-dimensional form; thus, the high-dimensional quantum state can be applied to complex calculations, including SQPC [33,34]. So far as we know, there is no relevant publication for the MQPC in size, which involves MQPC and SQPC into one.

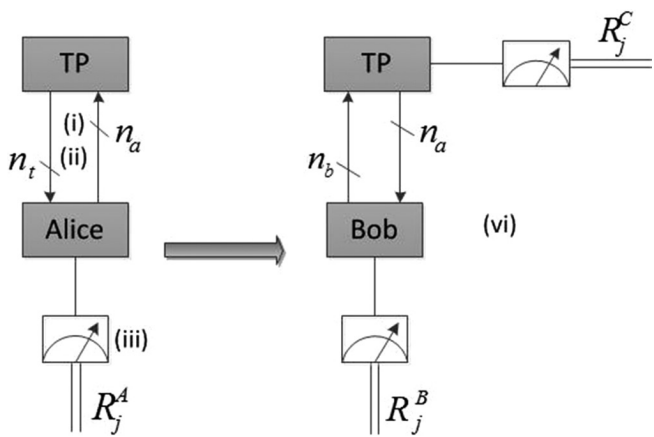

**Figure 3:** Process of the entanglement swapping QPC.

Being the foundation of quantum secure computation, QPC has extensively application prospect, such as quantum multi-party summation, quantum voting, and quantum auction. Recently, Chen *et al.* has proposed a multi-party quantum summation protocol [35] based on multi-particle GHZ states, which allows a group of mutually distrustful players to perform the summation computation. In order to satisfy the privacy condition, the quantum anonymous voting [36,37] was proposed to implement tasks related to secret voting and maintaining the anonymity of the voters. Quantum auction is another kind of application in commerce business, which includes three transactional types: The ascending-bid auction (English auction) [38], the descending-bid auction (Dutch auction), the first-price sealed-bid auction, and the second-price sealed-bid auction (Vickrey's action) [39,40]. Because quantum entanglement is not observable by others, the bidders who share the entangled states can arrange for correlations among their bids, thus non-winning bids are never revealed. Quantum English auction and quantum Vickrey's action have already been studied by some researchers. However, because Dutch action is non-commonly used in commerce, there is no relevant research up till now.

Up to date, QPC and other relevant issues have attracted more and more attentions of researchers, and also achieved some outcomes in theory. We can predict they will continue to be developed by leaps and bounds in the next few decades. Maybe, the following three directions will become the research mainstream in the future work,

- Improvement of existing quantum secure computation issues Nowadays, all kinds of QPC, quantum multi-party summation, quantum voting, and quantum auctions have been proposed, taking security and efficiency into consideration, there are probably many aspects to be improved
- Other complex QSMC Such computations, where secrecy is an issue, include joint data base computation, private and secure database access, joint signatures, joint decryption, and any other multi-party computation [41]
- Formalism analysis of quantum secure computation The security analysis of quantum secure computation is mostly based on the non-formalized analytical method (also called attack verification method). In the security analysis, some given attacks are adopted to assault the present protocols; thus, this kind of analytical method has some shortcomings: (a) The method lacks strictness, for it is impossible to adopt all kinds of attacks. (b) It can be proven to be secure against the known attacks, but cannot confirm whether there is no flaw in the whole protocol. As we know, the automated verification techniques [42] have made important achievements in the classical computation fields, which have already been extensively used to analyze the security of the protocols. Having made use of the present quantum formalism outcomes, such as quantum logic [43] and quantum program language [44], we can introduce the formalism analysis into the quantum secure computation. I think, it is a most worthwhile research topic.

## 6. Acknowledgment

This work is supported by China National Nature Science Foundation (Grant Nos. 61103235 and 61170321), A Project Funded by the Priority Academic Program Development of Jiangsu Higher Education Institutions, and Natural Science Foundation of Jiangsu Province, China (BK2010570).

## AUTHORS

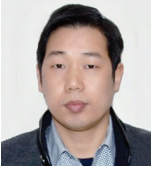

**Wenjie Liu** was born in 1979. He is an associate professor of Computer Sciences at Nanjing University of Information Science and Technology, China. He received his Bachelor (HZAU, China, 1997), Master (WHU, China, 2004), Ph.D. (SEU, 2011). His research interests include quantum secure multi-party computation, quantum secure communication, quantum algorithm, etc.

E-mail: wenjiel@163.com

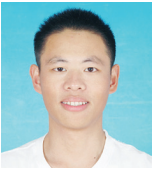

**Chao Liu** Received the B. A. degree in Network engineering from NUIST, China. He is currently working toward the Master's degree at the School of Computer and Software, NUIST. His research interests are quantum secure computation, quantum-inspired evolution algorithms, etc.

E-mail: ChaoLiu.Net@gmail.com

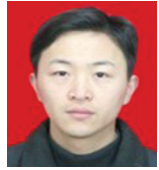

**Haibin Wang** is a lecturer of Computer Sciences at Nanjing University of Information Science and Technology, China. He received his Bachelor (NUIST, China, 1997), Master (NUIST, China, 2004), and is currently working toward the doctor's degree at the School of Computer and Software, NUIST. His research interests are Information security, quantum neuron network, etc.

E-mail: whb9741705@163.com

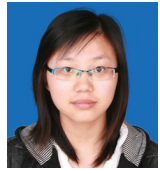

**Tingting Jia** Received the B.A. degree in Electrical and information engineering from NUIST, China. She is currently working toward the Master's degree at School of Computer and Software, NUIST. Her research interests are quantum teleportation, quantum remote state preparation, etc.

E-mail: sunnyting0521@126.com